\documentclass[a4paper,12pt]{article}
\pdfoutput=1

\usepackage{devanagari}

\usepackage{amsmath,amssymb,amsfonts, bm}
\usepackage{dsfont}
\usepackage{epsfig}
\usepackage{graphicx}
\usepackage{slashed}
\usepackage{color}
\usepackage{tipa}

\usepackage{caption}
\usepackage{subcaption}
\usepackage{slashed}

\usepackage{commath}
\usepackage{calc}

\newcommand{\be}{\begin{equation}}
\newcommand{\ee}{\end{equation}}

\newcommand{\bea}{\begin{eqnarray}}
\newcommand{\eea}{\end{eqnarray}}

\newcommand{\bfig}{\begin{figure}}
\newcommand{\efig}{\end{figure}}

\def \jhep{ {\bf JHEP}  }
\def \plb{ {\bf Phys. Lett. B} }

\usepackage{colortbl}

\definecolor{Gray}{gray}{0.95}
\definecolor{RGray}{gray}{0.85}
\definecolor{CGray}{gray}{0.92}

\usepackage{multicol}
\definecolor{tit}{rgb}{0.1,0.2,0.4}
\definecolor{blus}{cmyk}{1,1,0,0.6}
\definecolor{verde}{cmyk}{0.92,0,0.59,0.25}

\usepackage{color,hyperref}
\definecolor{darkblue}{rgb}{0.0,0.0,0.3}
\hypersetup{colorlinks,breaklinks,
            linkcolor=darkblue,urlcolor=darkblue,
            anchorcolor=darkblue,citecolor=darkblue}

\providecommand*\url[1]{\href{#1}{#1}}
\renewcommand*\url[1]{\href{#1}{\texttt{#1}}}

\newcommand{\D}{{\cal D}}
\newcommand{\U}{{\cal U}}

\newcommand{\M}{{\cal M}}

\makeatletter
\newcommand*{\rom}[1]{\expandafter\@slowromancap\romannumeral #1@}
\makeatother

\begin{document}

\allowdisplaybreaks
\vspace*{-2.5cm}
\begin{flushright}
{\small
IIT-BHU
}
\end{flushright}

\vspace{2cm}

\begin{center}
{\LARGE \bf \color{tit}
Solving the fermionic mass hierarchy of the standard model}\\[1cm]

{\large\bf Gauhar Abbas$^{a,b}$\footnote{email: gauhar.phy@iitbhu.ac.in}}  
\\[7mm]
{\it $^a$ } {\em Department of Physics, Indian Institute of Technology (BHU), Varanasi 221005, India}\\[3mm]
{\it $^b$ } {\em Theoretical Physics
Division, Physical Research Laboratory, Navrangpura, Ahmedabad 380
009, India}\\[3mm]

\vspace{1cm}
{\large\bf\color{blus} Abstract}
\begin{quote}
We show that  a  simultaneous  explanation for fermionic mass hierarchy among and within the fermionic families, quark-mixing, can be obtained  in an extension of the standard model, with real   singlet scalar fields, which is UV completed by vector-like fermions and a strongly interacting sector.
\end{quote}

\thispagestyle{empty}
\end{center}

\begin{quote}
{\large\noindent\color{blus} 
}

\end{quote}

\newpage
\setcounter{footnote}{0}

\section{Introduction}
In a recent interview published in CERN COURIER, Steven Weinberg was asked what single open question  he would like to see  answered in his lifetime, and Weinberg replied  that it is  only the mystery of  the observed pattern of quarks and leptons masses\cite{CC:13oct17}.

The fermionic mass hierarchy in the standard model (SM) is a complex problem, and can  be divided into three hierarchies.  The first is the mass hierarchy among fermionic families, i.e. $ m_\tau >> m_\mu >> m_e$, $ m_b >> m_s  >> m_d$ and $m_t  >> m_c >> m_u$.   The second hierarchy is within the family, i.e., $m_d > m_u$,   $m_c > > m_s$,  $m_t >> m_b$, and the third hierarchy resides in the quark-mixing angles, i.e. $\sin \theta_{12} >> \sin \theta_{23} >> \sin \theta_{13}$.

Serious efforts have been made to solve this problem within the framework of extended gauge or  flavour symmetries\cite{Froggatt:1978nt}-\cite{Camargo-Molina:2017kxd}.    For instance, a solution in Ref.\cite{Balakrishna:1987qd} depends on an extended nonabelian gauge sector and a number of scalar fields along with vector-like fermions.  The solution of Ref.\cite{Davidson:1993xn} also depends on an extended  nonabelian gauge sector, a global axial $U(1)$ symmetry and a number of scalar fields.  Other solutions are  based on extended gauge sectors such as Pati-Salam\cite{Pati:1974yy} and $SO(10)$  unification,  flavor nonabelian continuous and discrete symmetries.  Some solutions even employ  all  of these features, for instance see Refs.\cite{Ross:2002fb,Bazzocchi:2008rz,Hartmann:2014ppa}.   A  simultaneous solution of this problem within the gauge symmetry of the SM, and without any global nonabelian continuous or  discrete symmetries is a challenging problem.

In this paper, we present two different models to address this issue.  In the first model, a solution of the fermionic mass hierarchy among the three families is presented.  In the second model, we  present a fermionic and scalar extension of the SM which simultaneously have a  solution of the problem of the inter- as well as intra-family mass hierarchy of the SM as well as the small quark-mixing. 

The central idea of this work comes from the observation that in the SM the mass generation of the charged fermions of the  three families  is achieved by ad hoc insertion of the Yukawa Lagrangian of dimension 4.  This Lagrangian is indeed a selection in the sense that for recovering masses of fermions one can also choose the next operator which is of dimension 5 provided the Yukawa operator is forbidden by some symmetry. 

In this paper, we propose that masses of all fermions are generated through dimension-5 operators instead of the Yukawa Lagrangian.  This selection rule also put charged fermion masses on an equal footting with neutrino masses which are generated by the Weinberg operator within  the framework of the SM.  Having chosen this selection rule, we investigate the impact of this rule on the fermionic mass spectrum of the SM.

It turns out that for explaining the mass hierarchy among three families, atleast three real singlet scalar fields are required.  For explaining full pattern of the fermionic masses which includes inter- as well as intra-family mass hierarchy and quark-mixing, one needs exactly six real scalar fields.  This model has its  ultra-violet (UV) completion in the form of vector-like fermions and a strongly interacting sector. 
\section{Models}
We first discuss fermionic mass heirarchy among the three families in our first model.  For this purpose, we add three real singlet scalar fields {\dn k}$_1$, {\dn k}$_2$ and {\dn k}$_3$  to the SM whose transformations under $SU(3)_c \times SU(2)_L \times U(1)_Y$ are the following\footnote{Notation is taken from the Devanagari script. The consonant  letter ``\text{ {\dn k}}"(k\textschwa) is  pronounced as ``ka" in Kashmir. This notation is dedicated to Indian philosopher Kanada ({\dn kZAd}) who  introduced concepts of quantization of matter and fundamental particle in  the 6th Century BCE in India\cite{kanada}.  Kana  ({\dn kZ}) means particle in Sanskrit.}:
\begin{eqnarray}
\text{ {\dn k}} _1 :(1,1,0),~\text{{\dn k}} _2:(1,1,0),~ \text{{\dn k}} _3 :(1,1,0).
 \end{eqnarray} 

Furthermore,  we extend the symmetry of the SM by imposing discrete symmetries  $\mathcal{Z}_2$ and  $\mathcal{Z}_2^\prime$   on the right handed fermions of each family, scalar fields  $\text{ {\dn k}} _1$, $\text{ {\dn k}} _2$, and $\text{ {\dn k}} _3$ as shown in Table \ref{tab1}.

The Yukawa Lagrangian is now completely forbidden by $\mathcal{Z}_2$ and  $\mathcal{Z}_2^\prime$  symmetries.  The masses of fermions of three families are now recovered from dimension-5 operators given by the following equation:

\bea
\label{mass1}
{\mathcal{L}}_{mass} &=& \dfrac{1}{\Lambda} \Bigl[  \Gamma_1 \bar{\psi}_{L,q}^1  \varphi  d_R   \text{ {\dn k}} _1 +  \Gamma_2 \bar{\psi}_{L,q}^2    \varphi  s_R  \text{ {\dn k}} _2 +   \Gamma_3 \bar{\psi}_{L,q}^3  \varphi  b_R   \text{ {\dn k}} _3    \\ \nonumber
&+&    \Gamma_1^\prime  \bar{\psi}_{L,q}^1  \tilde{\varphi} u_R  \text{ {\dn k}} _1  
+ \Gamma_2^{ \prime} \bar{\psi^{2}_{L}}  \tilde{\varphi}  c_R  \text{ {\dn k}} _2 
+ \Gamma_3^\prime  \bar{\psi_L^3}  \tilde{\varphi} t_R   \text{ {\dn k}} _3   \\ \nonumber
&+&   \Gamma_e \bar{\psi}_{L,l}^1  \varphi  e_R   \text{ {\dn k}} _1 +  \Gamma_\mu \bar{\psi}^{2}_{L,l}    \varphi  \mu_R  \text{ {\dn k}} _2 +   \Gamma_\tau \bar{\psi}_{L,l}^3  \varphi  b_R   \text{ {\dn k}} _3  
\Bigr]   
+ \dfrac{c}{\Lambda} \bar{\psi}_{L,l}^c    \tilde{\varphi}^* \tilde{\varphi}^\dagger  \bar{\psi}_{L,l}
+  {\rm H.c.},
\eea
where $\bar{\psi}_{L,q}$ denotes the quark-doublet, $ \bar{\psi}_{L,l}$ is the leptonic doublet and superscripts denote the family number.

\begin{table}[h]
\begin{center}
\begin{tabular}{|c|c|c|}
  \hline
  Fields             &        $\mathcal{Z}_2$                    & $\mathcal{Z}_2^\prime$    \\
  \hline
  $u_R, d_R, e_R$                 &   +  &     -                                  \\
  $\text{ {\dn k}} _1$                        & +  &      -                                                 \\
   $c_R, s_R, \mu_R$                 &   -  &     -                                  \\
   $\text{ {\dn k}} _2$                        & -  &      -                                                \\
  $ t_R, b_R, \tau_R$     & -  &   +                                             \\
  $\text{ {\dn k}} _3$           & - &      +          \\
  \hline
     \end{tabular}
\end{center}
\caption{The charges of right-handed fermions of three families of the SM and singlet scalar fields under $\mathcal{Z}_2$ and  $\mathcal{Z}_2^\prime$  symmetries. Here, superscript denotes a family number.}
 \label{tab1}
\end{table} 

It is emphasized that mass hierarchy of fermionic families  is a repercussion of this model.  This is because  vacuum expectation values (VEVs) of the real singlet scalar fields are such that $ \langle \text{ {\dn k}} _3 \rangle >> \langle \text{ {\dn k}} _2 \rangle >> \langle \text{ {\dn k}} _1 \rangle $. This explains  why top quark is so heavy and electron is so light.   It should be noted that  this model has its own independent phenomenology which could be explored at the LHC as well as in low energy flavour physics. 

We discuss now a more powerful prediction of symmetries $\mathcal{Z}_2$ and $\mathcal{Z}_2^\prime$ along with a new added symmetry $\mathcal{Z}_2^{\prime \prime}$ in our second model.  It is natural to ask if a simultaneous explanation could be obtained for observed pattern of charged fermions among the three families as well as within the family.    The price to pay is to add six real singlet scalar fields  $\text{ {\dn k}} _i: (1,1,0)$ where $i=1-6$ which have charges under discrete symmetries $\mathcal{Z}_2$,  $\mathcal{Z}_2^\prime$ and $\mathcal{Z}_2^{\prime \prime}$ as given in  Table \ref{tab2}. 

 We note again that symmetries $\mathcal{Z}_2$,  $\mathcal{Z}_2^\prime$ and $\mathcal{Z}_2^{\prime \prime}$ forbid the Yukawa Lagrangian.   Masses of fermions  are again recovered by dimension-5 operators.  The mass Lagrangian now reads,

\bea
\label{mass2}
{\mathcal{L}}_{mass} &=& \dfrac{1}{\Lambda} \Bigl[  \Gamma_1 \bar{\psi_L^1}  \tilde{\varphi} u_R   \text{ {\dn k}} _1 +  \Gamma_2 \bar{\psi^{2}_{L}}     \tilde{\varphi} c_R  \text{ {\dn k}} _3 +   \Gamma_3 \bar{\psi_L^3}  \tilde{\varphi}  t_R   \text{ {\dn k}} _5   
+    \Gamma_1^\prime  \bar{\psi_L^1}  \varphi d_R  \text{ {\dn k}} _2  \\ \nonumber
&+& \Gamma_2^{ \prime} \bar{\psi^{2}_{L}}  \varphi   s_R  \text{ {\dn k}} _4
+ \Gamma_3^\prime  \bar{\psi_L^3}   \varphi  b_R   \text{ {\dn k}} _6  
+     \Gamma_e  \bar{\psi_L^1}  \varphi  e_R  \text{ {\dn k}} _2  + \Gamma_\mu \bar{\psi^{2}_{L}}   \varphi   \mu_R  \text{ {\dn k}} _4 \\ \nonumber
&+& \Gamma_\tau  \bar{\psi_L^3}  \varphi  \tau_R   \text{ {\dn k}} _6 \Bigr]  
+ \dfrac{c}{\Lambda} \bar{l_{L}^c}    \tilde{\varphi}^* \tilde{\varphi}^\dagger  l_L  
+  {\rm H.c.}.
\eea

Now, the whole observed mass pattern of charged fermions reveals itself aesthetically when six real singlet scalar fields $\text{ {\dn k}} _i$ acquire VEVs in such a way that  $ \langle \text{ {\dn k}} _2 \rangle > \langle \text{ {\dn k}} _1 \rangle $, $ \langle \text{ {\dn k}} _3 \rangle >> \langle \text{ {\dn k}} _4 \rangle $, $ \langle \text{ {\dn k}} _5 \rangle >> \langle \text{ {\dn k}} _6 \rangle $, $ \langle \text{ {\dn k}} _{6} \rangle >> \langle \text{ {\dn k}} _{4} \rangle >> \langle \text{ {\dn k}} _{2} \rangle $, and  $ \langle \text{ {\dn k}} _5 \rangle >> \langle \text{ {\dn k}} _3 \rangle >> \langle \text{ {\dn k}} _1 \rangle $.  Thus, this VEVs pattern explains  why $m_d > m_u$,  $m_c >> m_s$, $m_t >> m_b$, $ m_\tau >> m_\mu >> m_e$, $ m_b >> m_s  >> m_d$, and  $m_t  >> m_c >> m_u$.    We observe  that neutrino masses are generated by  the Weinberg operator which involves only the SM Higgs doublet field and can be recovered via celebrated seesaw mechanism.

\begin{table}[h]
\begin{center}
\begin{tabular}{|c|c|c|c|}
  \hline
  Fields             &        $\mathcal{Z}_2$                    & $\mathcal{Z}_2^\prime$   & $\mathcal{Z}_2^{\prime \prime}$      \\
  \hline
  $u_{R}$                 &   +  &     +    & -                                \\
  $\text{ {\dn k}} _1$                        & +  &      +  &-                                                 \\
   $d_{R}, e_R$                 &   -  &     -    & +                               \\
  $\text{ {\dn k}} _2$                        & -  &      -  &+                                                \\
   $c_{R}$                 &   +  &     -    & -                                \\
  $\text{ {\dn k}} _3$                        & +  &      -  &-                                                \\
   $s_{R}, \mu_R$                 &   +  &     -    & +                               \\
  $\text{ {\dn k}} _4$                        & +  &      - &+                                              \\
   $t_{R}$                 &   - &     +    & -                                \\
  $\text{ {\dn k}} _5$                       & -  &      +  &-                                                 \\
   $b_{R}, \tau_R$                 &   -  &     +    & +                              \\
  $\text{ {\dn k}} _6$                        & -  &      +  &+                                                \\
  \hline
     \end{tabular}
\end{center}
\caption{The charges of right-handed fermions of three families of the SM and singlet scalar fields under $\mathcal{Z}_2$, $\mathcal{Z}_2^\prime$ and $\mathcal{Z}_2^{\prime \prime}$ symmetries. We show flavour of right-handed fermion by superscript.}
 \label{tab2}
\end{table} 

\subsection{Ultraviolet completion}
\label{UVC}
A  UV completion of models described in tables \ref{tab1} and \ref{tab2}  can be achieved by introducing  one vector-like isosinglet up type quark, one vector-like isosinglet down type quark, and  one isosinglet vector-like charged lepton.  Their transformations  under $SU(3)_c \times SU(2)_L   \times U(1)_{Y} $ are given by,
\begin{eqnarray}
Q &=& U_{L,R} :(3,1,\dfrac{4}{3}),D_{L,R}:(3,1,-\dfrac{2}{3}),  
L =  E_{L,R}:(1,1,-2).
\end{eqnarray}

The mass Lagrangian for vector-like fermions is given by,
\begin{eqnarray}
\label{mass3}
\mathcal{L}_{V} &=& M_U \bar{U}_L U_R + M_{D} \bar{D}_L D_R~ 
+  M_E \bar{E}_L  E_R  + {\rm H.c.}.
\end{eqnarray}
The interactions of vector-like fermions with the SM  fermions, for instance for quarks, are given by,
\begin{eqnarray}
\label{mass4}
\mathcal{L} &=& Y_1  \bar{q}_L^1  \tilde{\varphi}  U_R  + Y_2  \bar{q}_L^2   \tilde{\varphi} U_R +  Y_3   \bar{q}_L^3  \tilde{\varphi}  U_R  \\ \nonumber
  &+& Y_1^\prime  \bar{q}_L^1 \varphi D_R  + Y_2^\prime  \bar{q}_L^2 \varphi D_R +  Y_3^\prime   \bar{q}_L^3 \varphi D_R 
  +  {\rm H.c}, 
\end{eqnarray}
where $q_L$ is quark doublet of the SM and superscript shows the family number.

The Lagrangian having interactions of singlet-scalar fields with the right-handed SM fermions can be written as,
\begin{eqnarray}
\label{mass4a}
\mathcal{L} &=&   C_1  \bar{U}_L  u_R \text{ {\dn k}} _{1}  
+ C_2 \bar{U}_L c_R \text{ {\dn k}} _{3}  + C_3    \bar{U}_L  t_R \text{ {\dn k}} _{5}  \\ \nonumber
&+& C_1^\prime  \bar{D}_L  d_R \text{ {\dn k}} _{2}  
+ C_2^\prime  \bar{D}_L s_R \text{ {\dn k}} _{4}  + C_3^\prime    \bar{D}_L  b_R \text{ {\dn k}} _{6} +  {\rm H.c}.
\end{eqnarray}
 We can write  similar Lagrangians  for leptons.  
 
Now we discuss what could be the scale of new physics(NP) entering in Eqs.(\ref{mass1})  and (\ref{mass2}).  The LHC searches  are indicating that the scale of vector-like fermions, which are NP in our model, should be greater than 900GeV\cite{atlas-cms-8,atlas-13}.  If we naively assume that the lightest VEV $v_1$ which couples to $u$ quark is of the order of 900GeV, for which there is no reason to assume,  the contribution of NP to the mass Lagrangian in Eq.\ref{mass2} is expected to be very small.  However, there is also contribution to masses from Eqs.(\ref{mass3}) to (\ref{mass4a}) within the renormalized theory.  Such contribution is nontrivial and recovers the physical mass as shown in Ref.\cite{Botella:2016ibj} where tree-level masses of light quarks are forbidden by a flavor symmetry.

We discuss now the diagonalization of mass matrices.  The mass matrix for down type quarks approximately reads,
\begin{equation}
\label{mmd}
\begin{array}{ll}
\M_\D = \left( \begin{array}{cccc}
\dfrac{\Gamma_{11}^d v v_2 }{2 M}&   \dfrac{\Gamma_{12}^d v v_4 }{2 M} & \dfrac{\Gamma_{13}^d v v_6 }{2 M} & \dfrac{1}{\sqrt{2}} v Y_1^d\\
  \dfrac{\Gamma_{21}^d  v v_2 }{2 M}&   \dfrac{\Gamma_{22}^d v v_4 }{2 M} & \dfrac{\Gamma_{23}^d v v_6 }{2 M}& \dfrac{1}{\sqrt{2}} v Y_2^d\\
  \dfrac{\Gamma_{31}^d v v_2 }{2 M}&   \dfrac{\Gamma_{32}^d v v_4 }{2 M} & \dfrac{\Gamma_{33}^d v v_6 }{2 M} &   \dfrac{1}{\sqrt{2}} v Y_3^d\\
 \dfrac{1}{\sqrt{2}} v_2 C_1^d   &     \dfrac{1}{\sqrt{2}} v_4 C_2^d     &    \dfrac{1}{\sqrt{2}} v_6 C_3^d        & M_{D}
\end{array} \right),
\end{array}
\end{equation} 
where $M$ is the scale of vector-like fermion and  VEVs of the  scalar fields are,
\bea
\langle \varphi \rangle &=& \dfrac{1}{\sqrt{2}}\begin{pmatrix} 0 \\ v \end{pmatrix}, ~\langle \text{ {\dn k}} _2 \rangle = v_2/\sqrt{2},~ \langle \text{ {\dn k}} _4 \rangle = v_4/\sqrt{2}, ~
 \langle \text{ {\dn k}} _6 \rangle = v_6/\sqrt{2}, \\ \nonumber
 \langle \text{ {\dn k}} _1 \rangle & =& v_1/\sqrt{2},~ \langle \text{ {\dn k}} _3 \rangle = v_3/\sqrt{2}, ~
 \langle \text{ {\dn k}} _5 \rangle = v_5/\sqrt{2}
\eea 
We recast mass matrix given in Eq.(\ref{mmd}) into the following form:
\begin{equation}
\label{MD2}
\begin{array}{ll}
\M_\D = \left( \begin{array}{cc}
m_d&   p \\
  X  & M_{D}
\end{array} \right) \,, \qquad 
\end{array}
\end{equation}
where  $m_d$ is the $3 \times 3$ SM fermionic block, $M_{D}$ is  $1 \times 1$  diagonal block. 

We  first diagonalize mass matrix $\M_\D$ through bi-unitary transformation,
\begin{equation}
\begin{array}{ll}
U^\dagger \M_\D V = \left( \begin{array}{cc}
\tilde{m}&   0 \\
  0 & \tilde{M}
\end{array} \right) \,, \qquad 
\end{array}
\end{equation}
where $\tilde{m} = \rm{diag} (m_d, m_s, m_b )$ and $\tilde{M} = M$ is the mass of the vector-like quark $D$.

The matrix $V$ diagonalizes $ \M_{\D}^\dagger \M_\D$ where $V$ is given by,
\begin{equation}
\begin{array}{ll}
V  = \left( \begin{array}{cc}
K_d &   R \\
  S & T
\end{array} \right).
\end{array}
\end{equation}
Using this, we obtain following relations:
\begin{subequations}
\bea
\label{eqa1}
(m_d^\dagger m_d + X^\dagger X ) K_d + ( m_d^\dagger p + X^\dagger M_{D} ) S &=& K_d \tilde{m}^2, \qquad  \\ 
\label{eqb1}
(m_d^\dagger m_d + X^\dagger X ) R + ( m_d^\dagger p + X^\dagger M_{D} )T &=& R \tilde{M}^2, \\
\label{eqc1}
(p^\dagger m_d + M_D^\dagger X) K_d  + (p^\dagger p + M_{D^{1}}^\dagger M_{D}) S &=& S \tilde{m}^2, \\
\label{eqd1}
(p^\dagger m_d + M_D^\dagger X)  R + (p^\dagger p + M_{D^{1}}^\dagger M_{D}) T &=& T \tilde{M}^2.
\eea
\end{subequations}
From Eq.(\ref{eqc1}), in the limit $\tilde{M}^2 >> \tilde{m}^2$ we obtain,
\be
S \simeq - (p^\dagger p + M_{D}^\dagger M_{D} )^{-1} (p^\dagger m_d + M_D^\dagger X) K_d.
\ee
From Eq.(\ref{eqa1}), we get,
\bea
K_d \mathcal{ H}_{eff} K_d^{-1}  = \tilde{m}^2,
\eea
where Hermitian squared matrix $\mathcal{ H}_{eff}$ is
\bea
\mathcal{ H}_{eff} &\simeq &  (m_d^\dagger m_d + X^\dagger X ) -  ( m_d^\dagger p + X^\dagger M_{D} ) 
  (p^\dagger p + M_{D}^\dagger M_{D})^{-1} (p^\dagger m_d + M_{D}^{\dagger} X).
\eea

\subsection{Quark mixing and masses of fermions}
The next challenge is to explain small quark-mixing along with the fermionic mass hierarchy among and within the three families.  This is again a formidable problem, and more difficult than the problem of fermionic mass hierarchy among and within the three families.   For this purpose, the mass matrix for down type quarks approximately reads,
\begin{equation}
\label{mmda}
\begin{array}{ll}
\M_\D = \left( \begin{array}{ccc}
\dfrac{\Gamma_{11}^d v v_2 }{2 M}&   \dfrac{\Gamma_{12}^d v v_4 }{2 M} & \dfrac{\Gamma_{13}^d v v_6 }{2 M} \\
  \dfrac{\Gamma_{21}^d  v v_2 }{2 M}&   \dfrac{\Gamma_{22}^d v v_4 }{2 M} & \dfrac{\Gamma_{23}^d v v_6 }{2 M} \\
  \dfrac{\Gamma_{31}^d v v_2 }{2 M}&   \dfrac{\Gamma_{32}^d v v_4 }{2 M} & \dfrac{\Gamma_{33}^d v v_6 }{2 M} 
\end{array} \right).
\end{array}
\end{equation}  
This matrix can be diagonalized through bi-unitary transformation
\begin{equation}
\begin{array}{ll}
U^\dagger_\D \M_\D V_\D = diag(m_d, m_s, m_b).
\end{array}
\end{equation}
In the limit $v, v_2, v_4 << v_6, M$, we obtain the following three mixing angles at leading order which completely parametrize the matrix $V_\D$\cite{Rasin:1998je}:
\bea
\tan \theta_{12}^d &\approx & \dfrac{ v_2}{ v_4 } C_{12}^d ,~
\tan \theta_{23}^d \approx  \dfrac{ v_4 }{ v_6 } C_{23}^d,~
\tan \theta_{13}^d \approx \dfrac{ v_2 }{ v_6  } C_{13}^d. 
\eea 
Similarly the matrix $V_\U$,  in the limit $v, v_1, v_3 << v_5, M$, can be derived for the mass matrix of up-type quarks which is parametrized by the following three mixing angles:
\bea
\tan \theta_{12}^u & \approx & \dfrac{ v_1}{ v_3} C_{12}^u,~
\tan \theta_{23}^u \approx  \dfrac{ v_3}{ v_5 } C_{23}^u,~
\tan \theta_{13}^u \approx   \dfrac{ v_1}{ v_5 } C_{13}^u. 
\eea 

The Cabibbo-Kobayashi-Maskawa  matrix is given by $V_{CKM} = V_\U^\dagger V_\D$, and the three mixing angles in the standard parameterization are,
\bea
\sin \theta_{12} &\approx &\dfrac{ v_2}{ v_4 } C_{12}^d ,~
\sin \theta_{23} \approx \dfrac{ v_4 }{ v_6 } C_{23}^d,~
\sin \theta_{13} \approx \dfrac{ v_2 }{ v_6  } C_{13}^d. 
\eea 
From the above results, it is remarkably obvious that we have obtained $\sin \theta_{12} >> \sin \theta_{23} >> \sin \theta_{13}$ in the limit $v_2 << v_4 << v_6$.

The masses of quarks at leading order in the limit  $v,v_1, v_2, v_3, v_4, v_6   << v_5, M$  given by\cite{Rasin:1998je}
\bea
m_u &\approx & \dfrac{ v v_1}{ 2 M}  y_u,~m_c \approx \dfrac{ v v_3}{ 2 M}  y_c, ~m_t \approx \dfrac{ v v_5}{ 2 M}  y_t,\\ \nonumber
m_d &\approx & \dfrac{ v v_2}{ 2 M}  y_d,~m_s \approx \dfrac{ v v_4}{ 2 M}  y_s,~m_b \approx \dfrac{ v v_6}{ 2 M}  y_b,
\eea 
where  we have assumed that couplings lie in the range $1\leq  y< 4 \pi$.  Thus, we observe that the VEVs of singlet scalar fields control everything from masses to mixing angles in this model.

For the sake of completeness, we present an example to show how well we can reproduce fermionic masses.  For this purpose, we choose $M= 1 \text{TeV}$ and  the following values of the quark masses at $\mu=m_t (m_t)$\cite{Xing:2007fb}
\bea
m_u &= & 1.22^{+ 0.48}_{-0.40} \text{MeV},~m_c = 0.59 \pm 0.08 \text{GeV}, ~m_t = 162.9 \pm 2.8 \text{GeV},\\ \nonumber
m_d &= &  2.76^{+ 1.19}_{-1.14} \text{MeV},~m_s = 52 \pm 15 \text{MeV},~m_b = 2.75 \pm 0.09 \text{GeV},
\eea 

We obtain a good fit for the following values of the fermionic couplings and VEVs of singlet scalar fields
\bea
v_1 &= & 0.009  \text{GeV},~v_2 = 0.02 \text{GeV}, ~v_3 = 4.8 \text{GeV},\\ \nonumber
v_4 &= & 0.092  \text{GeV},~v_5 =  1.33 \text{TeV},~v_6 = 6 \text{GeV}, \\ \nonumber
y_u &=& 1, ~ y_d=1,~ y_c = 1,~ y_s= 1,~ y_t= 1,~ y_b=3.7.
\eea 

Using the above values of the VEVs of the singlet scalar fields, we can compute the quark-mixing angles.  Assuming $C_{12,23,13}^d =1 $, we obtain,
\bea
\sin \theta_{12} &\approx &0.229,~
\sin \theta_{23} \approx 0.042,~
\sin \theta_{13} \approx 0.0033,
\eea
which is in agreement with experimental data.

The masses of charged leptons now can be predicted using the values of the VEVs of singlet scalar fields.  The approximate numerical values of the couplings required are,
\bea
y_e &\approx & 0.2, ~ y_\mu \approx  9.3,~ y_\tau = 2.44,
\eea 

As commented earlier, the neutrino masses can be recovered  through seesaw mechanism.  There are two ways to implement it.  The first method is to use TeV scale seesaw mechanisms discussed, for instance in Ref. \cite{Chen:2011de}.  The second solution may be to assume that new physics scale entering in the Weinberg operator is different from that enters in the dimension-5 operators used for charged fermion masses.  The main aim of this work is to investigate the charged fermion masses, and a  detailed study of neutrino masses is beyond the scope of this paper.

\subsection{Scalar potential}
We discuss now  the scalar potential in this model.   The most general scalar sector allowed by $\mathcal{Z}_2$, $\mathcal{Z}_2^\prime$ and $\mathcal{Z}_2^{\prime \prime}$ symmetries can be written as,
\bea 
V &=& \mu \varphi^\dagger \varphi + \lambda (\varphi^\dagger \varphi)^2  + \sum_{i=1}^6  \mu_i \text{ {\dn k}} _i^2  +   \varphi^\dagger \varphi   \sum_{i=1}^6  \lambda_i \text{ {\dn k}} _i^2 +   \sum_{i j}^6  \lambda_{ij}  \text{ {\dn k}}_i^2   \text{ {\dn k}}_j^2.  
\label{SP}
\eea

The masses of physical scalars can be recovered from the mass matrix of the scalar sector which is given as,
\begin{equation}
\M^2=\left(
\begin{array}{ccccccc}
 \dfrac{\lambda_{11} v_1^2}{2} & \dfrac{\lambda_{12} v_1 v_2}{2} & \dfrac{\lambda_{14} v_1 v_4}{2} & \dfrac{\lambda_{13} v_1 v_3}{2}  &   \dfrac{\lambda_{16} v_1 v_6}{2} &  \dfrac{\lambda_{1} v_1 v}{2}
 & \dfrac{\lambda_{16} v_1 v_5}{2} \\
 \dfrac{\lambda_{21} v_2 v_1}{2}  &  \dfrac{\lambda_{22} v_2^2}{2}   &
   \dfrac{\lambda_{24} v_2 v_4}{2}  &  \dfrac{\lambda_{23} v_2 v_3}{2}  &  \dfrac{\lambda_{26} v_2 v_6}{2}  &  \dfrac{\lambda_{2} v_2 v}{2}  &  \dfrac{\lambda_{25} v_2 v_5}{2}  \\
 \dfrac{\lambda_{41} v_4 v_1}{2}  &  \dfrac{\lambda_{42} v_4 v_2}{2}   &
   \dfrac{\lambda_{44} v_4^2 }{2}  &  \dfrac{\lambda_{43} v_4 v_3}{2}  &  \dfrac{\lambda_{46} v_4 v_6}{2}  &  \dfrac{\lambda_{4} v_4 v}{2}  &  \dfrac{\lambda_{45} v_4 v_5}{2}  \\
 \dfrac{\lambda_{31} v_3 v_1}{2}  &  \dfrac{\lambda_{32} v_3 v_2}{2}   &
   \dfrac{\lambda_{34} v_3 v_4 }{2}  &  \dfrac{\lambda_{33} v_3^2 }{2}  &  \dfrac{\lambda_{36} v_3 v_6}{2}  &  \dfrac{\lambda_{3} v_3 v}{2}  &  \dfrac{\lambda_{35} v_3 v_5}{2}  \\
 \dfrac{\lambda_{61} v_6 v_1}{2}  &  \dfrac{\lambda_{62} v_6 v_2}{2}   &
   \dfrac{\lambda_{64} v_6 v_4 }{2}  &  \dfrac{\lambda_{63} v_6 v_3 }{2}  &  \dfrac{\lambda_{66} v_6 v_6}{2}  &  \dfrac{\lambda_{6} v_6 v}{2}  &  \dfrac{\lambda_{65} v_6 v_5}{2}  \\
 \dfrac{\lambda_{1} v v_1}{2}  &  \dfrac{\lambda_{2} v v_2}{2}   &
   \dfrac{\lambda_{4} v v_4 }{2}  &  \dfrac{\lambda_{3} v v_3 }{2}  &  \dfrac{\lambda_{6} v v_6}{2}  &  2 \lambda v^2  &  \dfrac{\lambda_{5} v v_5}{2}  \\
    \dfrac{\lambda_{51} v_5 v_1}{2}  &  \dfrac{\lambda_{52} v_5 v_2}{2}   &
   \dfrac{\lambda_{54} v_5 v_4 }{2}  &  \dfrac{\lambda_{53} v_5 v_3 }{2}  &  \dfrac{\lambda_{56} v_5 v_6}{2}  &  \dfrac{\lambda_{5} v_5 v}{2}  &  \dfrac{\lambda_{55} v_5^2}{2}  \\
\end{array}
\right).
\end{equation} 

Since $v_{1,2,4} << v_{3,5,6}, v$, this matrix can be written in the following form:
\begin{equation}
\begin{array}{ll}
\M^2 = \left( \begin{array}{cc}
m_l&   S^T \\
  S & M_{H}
\end{array} \right) \,, \qquad 
\end{array}
\end{equation}
where  $m_l$ is the $3 \times 3$  block and  $M_{H}$ is  $4 \times 4$  diagonal block. 

We  block-diagonalize mass matrix $\M^2$ through orthogonal transformation,
\begin{equation}
\begin{array}{ll}
W^T \M^2 W = \left( \begin{array}{cc}
M_{light}&   0 \\
  0 & M_{heavy}
\end{array} \right)
\end{array}.
\end{equation}
The approximate form of the matrix $W$ is given by\cite{Grimus:2000vj},
\begin{equation}
\begin{array}{ll}
W = \left( \begin{array}{cc}
1 &   B \\
  -B^\dagger & 1
\end{array} \right),
\end{array}
\end{equation}
where $B \approx (1+   Z^\dagger Z )^{-1/2}  Z^\dagger$, $Z= M_H^{-1} S$, $M_{light} \approx m_l - S^T M_H^{-1} S$, and $M_{heavy} = M_H$.

The leading contribution to  mass square eigenvalues of the matrix  $M_{light}$ ignoring the contribution from the term $ S^T M_H^{-1} S$ is given by, 
\bea
m_{S_1}^2 &\approx&  \frac{ v_1^2 (\lambda_{11} (\lambda_{24}^2-4 \lambda_{22} \lambda_{44})+\lambda_{12}^2 \lambda_{44}-\lambda_{12} \lambda_{14} \lambda_{24}+\lambda_{14}^2 \lambda_{22})}{2 (\lambda_{24}^2-4 \lambda_{22} \lambda_{44})}   + \mathcal{O} (\dfrac{v_1^3}{v_4^3}) \\ \nonumber
m_{S_2}^2 &\approx& \dfrac{1}{4} \Bigl(-\sqrt{\lambda_{22}^2 v_2^4+v_2^2 v_4^2 (\lambda_{24}^2-2 \lambda_{22} \lambda_{44})+\lambda_{44}^2 v_4^4}   + \lambda_{22} v_2^2 + \lambda_{44} v_4^2  \Bigr) + \mathcal{O} (\dfrac{v_1^2}{v_4^2}) \\ \nonumber
m_{S_4}^2 &\approx& \dfrac{1}{4} \Bigl(\sqrt{\lambda_{22}^2 v_2^4+v_2^2 v_4^2 (\lambda_{24}^2-2 \lambda_{22} \lambda_{44})+\lambda_{44}^2 v_4^4}   + \lambda_{22} v_2^2 + \lambda_{44} v_4^2  \Bigr) + \mathcal{O} (\dfrac{v_1^2}{v_4^2}). 
\eea
The matrix $M_{heavy} $ can be further decomposed into two blocks of dimensions $2\times 2$ by noting $v_{3,6} << v, v_5$.  The leading contribution to mass square eigenvalues again is
\bea
m_{S_3}^2 &\approx&  \dfrac{1}{4} \Bigl(-\sqrt{\lambda_{33}^2 v_3^4+v_3^2 v_6^2 (\lambda_{36}^2-2 \lambda_{33} \lambda_{66})+\lambda_{66}^2 v_6^4}   + \lambda_{33} v_3^2 + \lambda_{66} v_6^2  \Bigr)   \\ \nonumber
m_{S_6}^2 &\approx&  \dfrac{1}{4} \Bigl(\sqrt{\lambda_{33}^2 v_3^4+v_3^2 v_6^2 (\lambda_{36}^2-2 \lambda_{33} \lambda_{66})+\lambda_{66}^2 v_6^4}   + \lambda_{33} v_3^2 + \lambda_{66} v_6^2  \Bigr)   \\ \nonumber
m_{H}^2 &\approx& \dfrac{1}{4} \Bigl(- \sqrt{\lambda_{55}^2  v_5^4+ 4 v^2 v_5^2 (\lambda_5^2-2 \lambda \lambda_{55})+ 16 \lambda^2 v^4}   + \lambda_{55} v_5^2 + 4 \lambda  v^2  \Bigr)  \\ \nonumber
m_{S_5}^2 &\approx& \dfrac{1}{4} \Bigl(\sqrt{\lambda_{55}^2  v_5^4+ 4 v^2 v_5^2 (\lambda_5^2-2 \lambda \lambda_{55})+ 16 \lambda^2 v^4}   + \lambda_{55} v_5^2 + 4 \lambda  v^2  \Bigr), 
\eea
where $m_H$ is the SM  like Higgs boson.

The potential contains many terms whose behaviour should be such that it remains consistent with the hierarchies of VEVs.  It needs a thorough numerical investigation which is a subject of a future study.  However, we can note from above results some main conditions which should be obeyed to achieve positive and hierarchical mass square eigenvalues.  These are,
\begin{eqnarray}
&& \lambda_{24}^2-4 \lambda_{22} \lambda_{44} >0, \lambda_{36}^2-2 \lambda_{33} \lambda_{66})>0, \lambda_5^2-2 \lambda \lambda_{55}>0, \lambda_{55} > \lambda_{33,66}>\lambda_{22,44} >\lambda_{11}, \\ \nonumber
&&  \lambda_{11} (\lambda_{24}^2-4 \lambda_{22} \lambda_{44})+\lambda_{12}^2 \lambda_{44} +\lambda_{14}^2 \lambda_{22} > \lambda_{12} \lambda_{14} \lambda_{24}.
\end{eqnarray}

We note that there are VEVs hierarchies in the model.  Their protection against quantum corrections can come from a larger underlying theory such as supersymmetry or  a theory where the scalar particles of the model are composite coming from a strongly interacting sector.

It is further noted that the model can predict five scalar particles lighter than the discovered SM like Higgs boson.  This is more elaborated in the next section.

\section{Phenomenological consequences}
We discuss now some immediate phenomenological implications.  It is indeed interesting to note that the  CMS collaboration  recently has reported an excess of events above the background in the dimuon mass spectrum at 8 and 13 TeV corresponding to integrated luminosity of 19.7 and 35.9 fb$^{-1}$, respectively\cite{Sirunyan:2018wim}.  The excess at 8 TeV is observed with a dimuon mass near 28 GeV in the central and the forward region corresponding to local significances of 4.2 and 2.9 standard deviations, respectively.  The analysis at 13 TeV has a mild excess only in the central region corresponding to a local significance of 2.0 standard deviations, while the forward region has a deficit with a local significance of 1.4 standard deviations.

The CMS collaboration has concluded that the two-Higgs-doublet models and next-to-minimal supersymmetric standard model (SM) cannot account for these  excesses after including theoretical and experimental constraints\cite{Sirunyan:2018wim}.  Furthermore, the CMS collaboration has concluded that the 13 TeV data cannot nullify the excess observed at the 8 TeV due to the lack of a realistic signal model. 

It is remarkable that a quite similar excess around 30 GeV of dimuon mass, having a local significance of 5 standard deviations,  is also reported in the re-analysed data of  the ALEPH experiment at LEP\cite{Heister:2016stz}.  This almost forgotten observation has phenomenal similarity with that is observed by the  CMS collaboration.    For a clear and transparent picture, we show the CMS and the ALEPH data in table \ref{tab4}.

\begin{table}[h]
\begin{center}
\begin{tabular}{|c|c|c|c|}
  \hline
  Experiments             &      Mass  (GeV)                 & Events  & Width (GeV)     \\
  \hline
  CMS                &   28.40 $\pm$ 0.60  &     22.0 $\pm$ 7.6   & 1.9$\pm$ 1.3                             \\
    \hline
 ALEPH                       & 30.40 $\pm$ 0.46  &      32.31 $\pm$ 10.87  & 1.78 $\pm$ 1.14                                               \\
  \hline
     \end{tabular}
\end{center}
\caption{}
 \label{tab4}
\end{table} 

Therefore, if the excesses observed in the CMS and ALEPH data are valid,  without reservation, it is a challenge to propose their theoretical origin.  We should note that two different and independent excesses from the CMS and the ALEPH data could be due to one scalar boson, or they can originate from the two almost degenerate scalar bosons.

We note that  the observed excesses at the CMS and the ALEPH experiments may correspond to the scalar bosons  either two of $m_{S_1}$, $m_{S_2}$ and $m_{S_4}$ through the processes $gg \rightarrow H \rightarrow S_i S_i \rightarrow \bar{b} b \mu^+\mu^-$ where $i=1,2,4$.  

Furthermore, the CMS collaboration has also observed an excess of mass 95.3 GeV for a low-mass scalar particles decaying to two photons in  data samples corresponding to integrated luminosity of 19.7 and 35.9 $\text{fb}^{-1}$ for the centre-of-mass energies of 8 and 13 TeV, respectively\cite{Sirunyan:2018aui}.  The local significance of this excess is 2.8 standard deviations. It should be noted that the scalar states $m_{S_3}$ or $m_{S_6}$ may account for this excess through the process $gg \rightarrow S_{3,6} \rightarrow \gamma \gamma$.

\section{An origin in a strongly interacting sector}

The main criticism of this work is definitely the presence of six hierarchical VEVs corresponding to the six singlet scalar fields.  This is equivalent to replacing the hierarchy of the Yukawa couplings by the hierarchy of the six VEVs corresponding to the six singlet scalar fields.  We note that it is not theoretically guaranteed that a law must exists to explain the hierarchy of the Yukawa couplings.  It could be that  nature has chosen the  hierarchical Yukawa couplings  to provide masses to fermions, and there exits nothing that can replace the smallness of these Yukawa couplings.  If this is so, a worth exploring and equivalent scenario is the hierarchy of VEVs which originates from the dynamics, rather than the static hierarchy of the Yukawa couplings.  In this sprite, the model presented in this work should be  considered.

Having said above, we argue that an underlying theory which could provide an explanation for six singlet scalar fields and their hierarchical VEV pattern can originate from a strongly interacting sector where the SM Higgs doublet and singlet scalar fields are bound states\cite{Weinberg:1975gm}-\cite{Agashe:2004rs}.  Such models are quite attractive since they can provide a solution of  the hierarchy problem\cite{Kaplan:1983fs,Kaplan:1983sm,Agashe:2004rs}. The present LHC status of these models is discussed in Ref. \cite{Sanz:2017tco,Belyaev:2018jse}.

Now we discuss a realization of such  UV completions for the model presented in this work at qualitative level. We assume a strongly-interacting sector which is conceptually identical to QCD like confining theory.  It is easy to discuss the fermionic masses in technicolour models\cite{Weinberg:1975gm,Susskind:1978ms,Dimopoulos:1979es}.  The argument can be extended to composite Higgs models\cite{Kaplan:1983fs,Kaplan:1983sm,Agashe:2004rs}.

We consider extended technicolour models where we may have a full symmetry, for instance, $SU(3)_c \times SU(2)_L \times U(1)_Y \times  SU(N)_{\text{ETC}}  \times SU(N_7)_{\text{ETC}} \times \mathcal{Z}_2 \times \mathcal{Z}_2^\prime \times \mathcal{Z}_2^{\prime \prime}$.  We note that ETC stands for extended technicolor.   

Now the extended technicolor symmetry is broken in six stages to technicolor symmetry in the following way:
\begin{eqnarray}
SU(N_7)_{\text{ETC}}  &\xrightarrow[\mu_6]{} &  SU(N_6)_{\text{ETC}} \xrightarrow[\mu_5]{}   SU(N_5)_{\text{ETC}} \xrightarrow[\mu_4]{}   SU(N_4)_{\text{ETC}} \xrightarrow[\mu_3]{}   SU(N_3)_{\text{ETC}} \\ \nonumber
& \xrightarrow[\mu_2]{} &   SU(N_2)_{\text{ETC}}  
\xrightarrow[\mu_1]{}   SU(N_1)_{\text{TC}},    
\end{eqnarray}
where $N_7>N_6>N_5>N_4>N_3>N_2>N_1 $ and $\mu_i$ ($i=1-6$) are breaking scales of symmetries.

The masses of fermions are recovered through quark condensates which are formed by the following  fermions:
\begin{eqnarray}
T_{L}={\begin{pmatrix} U \\ D \end{pmatrix}}_{L}:(1,2,\dfrac{1}{3},N,1), T_R= U_{R}:(1,1,\dfrac{4}{3},N,1),D_{R}:(1,1,-\dfrac{2}{3},N,1).
\end{eqnarray} 
Besides above for each SM quark, we add a vector-like quark such that we have the following quarks:
\begin{eqnarray}
Q &=& U_{L,R} :(1,1,\dfrac{4}{3},1,N^{1-7}),D_{L,R}:(1,1,-\dfrac{2}{3},1, N^{1-7}).
\end{eqnarray}

The typical form of new interactions involving SM  and vector-like quarks can be written as,
\begin{eqnarray}
\mathcal{L}_{\text{strong}} &= & \sum_{i=1}^6 \dfrac{g_{\text{ETC}}}{\Lambda} \dfrac{\left( \bar{T}_L  T_R\right)}{\mu^{2}}  \dfrac{\left( \bar{Q}_{L} Q_R   \right)}{\mu_i^2} \left( \bar{\psi}_L \psi_R  \right),
\end{eqnarray}
where  $\psi$ represents the SM fermions. $g_{\text{ETC}}$ is the coupling constant  of the extended technicolour group.

After spontaneous breaking of the global symmetry of the strongly interacting sector by new quarks condensate, i.e $<\left( \bar{T}_L  T_R\right) > , <\left( \bar{Q}_{L} Q_R   \right)>\neq 0$ , the masses of the SM fermions are given by
\begin{eqnarray}
m_\psi^i &=&   \dfrac{g_{\text{ETC}}}{\Lambda} \dfrac{|<\left( \bar{T}_L  T_R\right) >|}{\mu^2} ~  \dfrac{|<\left( \bar{Q}_{L} Q_R   \right)>|}{\mu_i^2},
\end{eqnarray}
where $Q = U \  \text{or} \ D$.  We must note that the Yukawa-like effective couplings are forbidden by already imposed three $\mathcal{Z}_2$ symmetries in table \ref{tab2}. 

Thus, the hierarchies of the six singlet scalar VEVs can be explained through the operators of the form $ \dfrac{|<\left( \bar{Q}_{L} Q_R   \right)>|}{\mu_i^2}= \langle \text{ {\dn k}} _i \rangle$ where $i=1-6$.  We stress that the six singlet VEVS depend on a factor of $1/\mu_i^2$.  This is the reason that they are  hierarchical.  This argument can also be extended to other popular composite Higgs models\cite{Kaplan:1983fs,Kaplan:1983sm,Agashe:2004rs}.

\section{Discussion and conclusion}
We conclude by summarizing what we have achieved in the extension of the SM presented in this work.  We have presented a selection rule  which  employs the dimension-5 operators to generate masses of all fermions of the SM. This results  an  extension of  the SM in fermionic and scalar sectors.  On the other side, symmetry is extended by discrete symmetries $\mathcal{Z}_2$, $\mathcal{Z}_2^\prime$ and $\mathcal{Z}_2^{\prime \prime}$ leading to two class of models.  The dimension-5 operators  may be UV completed by vector-like fermions  having interesting phenomenology scenarios\cite{Alok:2015iha}-\cite{Abbas:2017jkx}.  This model allows us to explain simultaneously fermionic mass hierarchy among and within the  three families, quark-mixing. One of the main features  of this work could be prediction of light scalars which can be probed by the LHC.   We wish to do a thorough phenomenological investigation of this model in future.

\end{document}